# Nuclear Data Needs for Microcalorimetry and Non-destructive Assay

Geon-Bo Kim[1]*, Begona Aranguren-Barrado[2], Shamsuzzoha Basunia[3], Daniel Becker[4,5], Lee Bernstein[3,6], Mark Croce[7], Joel Ullom[4,5], and Andrew Voyles[3,6]

*[1]Lawrence Livermore National Laboratory, Livermore, CA 94550, USA*
*[2]Department of Energy National Nuclear Security Administration, Washington DC 20585, USA*
*[3]Lawrence Berkeley National Laboratory, Berkeley, CA 94720, USA*
*[4]University of Colorado, Boulder, CO 80309, USA*
*[5]National Institute of Standards and Technology, Boulder, CO 80305, USA*
*[6]University of California – Berkeley Dept. of Nuclear Engineering, Berkeley, CA 94720, USA*
*[7]Los Alamos National Laboratory, Los Alamos, NM 87545, USA*
**Email: kim90@llnl.gov*

**Abstract:** Cryogenic microcalorimeters are state-of-the-art radiation detectors using superconducting and quantum technologies. They can resolve complex X-ray and low-energy γ-ray spectra with ultra-high energy resolution on the order of 10 eV at 100 keV, enabling high-precision non-destructive assay (NDA) analysis of nuclear materials containing uranium, plutonium and other actinides. With significant technical advancements in microcalorimetry technology, microcalorimeters are now deployable to end-users such as the International Atomic Energy Agency (IAEA) for improved NDA. However, the accuracy of microcalorimetry analysis can be limited by nuclear data. There are several cases that the current nuclear data obtained by conventional radiation detector technologies is not sufficient to support microcalorimetry analysis. To address the growing need for improved nuclear data in microcalorimetry, the U.S. Department of Energy Office of International Nuclear Safeguards hosted a workshop on Microcalorimetry and Nuclear Data (MiND) in June 2023. Microcalorimetry experts and users, and nuclear structure evaluators and managers, and program sponsors attended the workshop with the main objective of identifying a roadmap for priority nuclear data, stakeholders, partnerships, and opportunities. This paper summarizes the outcome of the MiND workshop, including the priority list of nuclear data for microcalorimetry and a multi-laboratory measurement campaign to improve such nuclear data.

## 1. Introduction

Microcalorimeters are radiation detectors based on precise thermometry at cryogenic temperatures of $T < 0.1$ K. Temperature increases of radiation absorbers are measured by extremely precise cryogenic thermometers. Magnetic Micro-Calorimeters (MMCs) [1] and Transition Edge Sensors (TESs) [2] are types of thermometers utilizing superconducting and quantum sensing technologies for improved thermometry precision. Unlike semiconductor or scintillator detectors, energy resolution $\Delta E$ of microcalorimeters is not limited by the Fano factor [3], instead it is limited by thermodynamic fluctuation noise that is described by the equation

$$\Delta E = \sqrt{k_B C T^2},$$

where $k_B$ is the Boltzmann constant, $C$ is the heat capacity of the detector, and $T$ is the temperature of the detector. For this reason, microcalorimeters are made to be small (micrometers to millimeters scale) and operated at cryogenic temperatures for lower heat capacities and higher energy resolution. These two factors, small detector volume and cryocooling requirement, are major challenges for utilizing microcalorimeters in practical applications.

There have been significant technological advancements of refrigeration instruments, especially with respect to user-friendly operation and automation. Modern refrigerators including dilution refrigerators (DRs) and adiabatic demagnetization refrigerators (ADRs) feature fully automatic operation with user friendly software, short cooldown time, and compact footprint enabling table-top operation. Operating these cryogenic refrigerators does not require a cryogenic expert and can be performed by technicians after several days of training. The challenge of small detector volume and low detection efficiency is mitigated by development of microwave multiplexing (μMUX) technology [4]. Hundreds of microcalorimeter pixels can be read out via a few readout channels and electronics with this technique, enabling operation of microcalorimeters at relatively higher detection efficiencies with centimeter-scale detection areas [5]. Based on these successful advancements, TES microcalorimeter instruments for gamma-ray spectrometers are being developed and deployed at user facilities [5].

Microcalorimeters are promising alternatives for future non-destructive assay (NDA) of nuclear materials. The primary energy range for analysis is at a few hundred kiloelectronvolts (keV), where a large number of X-rays and weak gamma-ray lines are present, making analysis challenging with conventional gamma-ray spectrometers with ~500 eV FWHM (full-width-half-maximum) energy resolution at these energies. This NDA with low energy gamma-rays could therefore significantly benefit from microcalorimeters with excellent energy resolution that can resolve complex gamma spectra and enable the detection of weak emission lines above the Compton background. Figure 1 shows two gamma-ray spectra of U-233 and mixed Pu sources measured by MMC and TES gamma-ray spectrometers. While gamma-ray lines are overlapping in the HPGe spectra, microcalorimeters with superior energy resolution can resolve most of the emission lines.

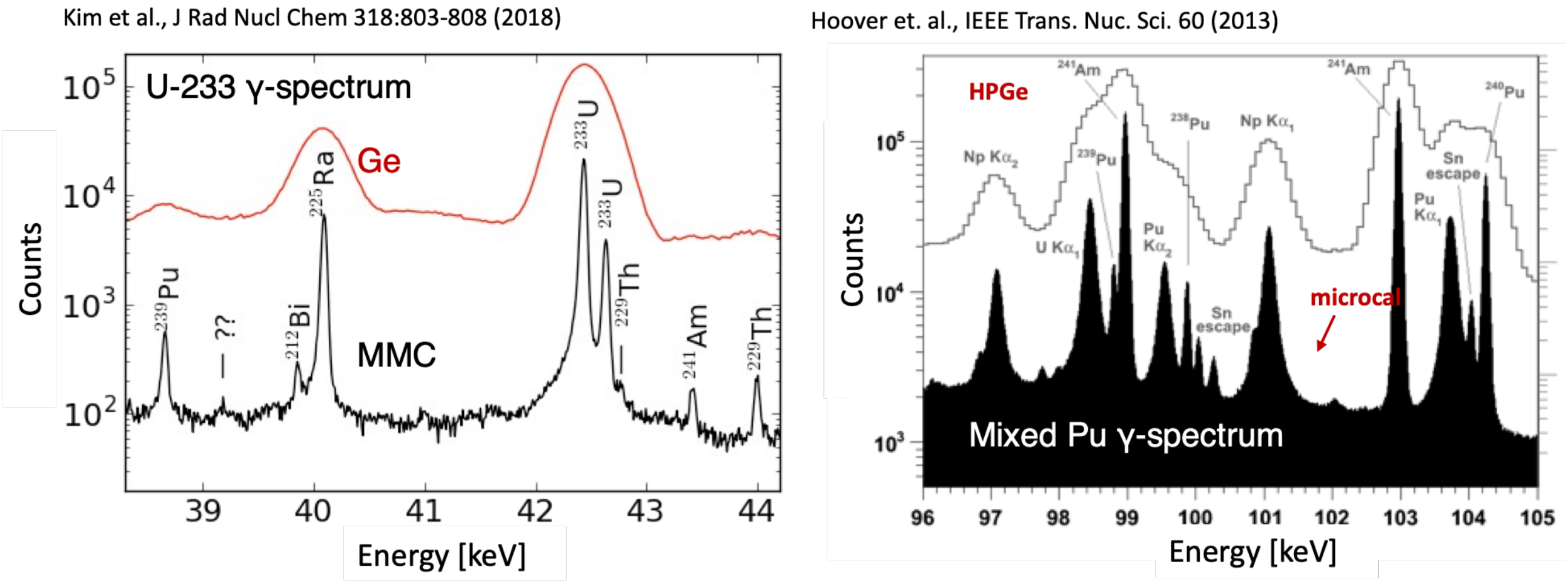

*Figure 1. (Left) U-233 gamma-ray spectra measured by HPGe and MMC gamma-ray spectrometers. (Right) Pu spectra measured by HPGe and TES gamma-ray spectrometers. TES microcalorimeter resolve complex X-ray and gamma-ray structures in the region of interest, and it requires accurate X-ray and gamma-ray intensity information to obtain accurate isotopic composition of Pu materials.*

Composition of nuclear materials can be obtained by extracting gamma and X-ray peak areas and converting them to activities of isotopes. There are a number of factors affecting uncertainties of this analysis, such as line energies, intensities, half-lives, and detection efficiency. By resolving individual gamma-ray and X-ray lines, microcalorimeters significantly reduce uncertainties arising from overlapping peaks. However, uncertainties from intensities remain unchanged, limiting overall accuracy of microcalorimetry analysis. Yoho et al. [6] reported uncertainty budget of microcalorimetry analysis and concluded that intensities are the dominant uncertainty factor for microcalorimetry Pu analysis.

Most of current nuclear decay data is obtained by conventional radiation detector technologies and with careful evaluation procedures. This data is sufficiently accurate for current technologies. However, it may not satisfy the new technology such as microcalorimeters with superior energy resolution and precision. For example, figure 2 shows gamma-ray spectra obtained by the HPGe detectors and microcalorimeters. Microcalorimeters with narrower peak widths can resolve small and weak gamma-ray lines that were not clearly observed in HPGe detectors. In this energy range, there are three major gamma-ray lines from U-233 decay. There are three gamma-ray peaks listed in the ENSDF (Nudat 3.0) library [7]. They are at 96.22(3) keV, 97.1346(3) keV, and 97.37(4) keV, with emission probabilities of 0.00170(9)%, 0.0203(10)%, and 0.0020(6)%, respectively. However, the third line at 97.37(4) keV was not observed despite sufficient statistical sensitivity. Because this triplet can provide key signatures for identifying and quantifying U-233, improved gamma-ray data are recommended.

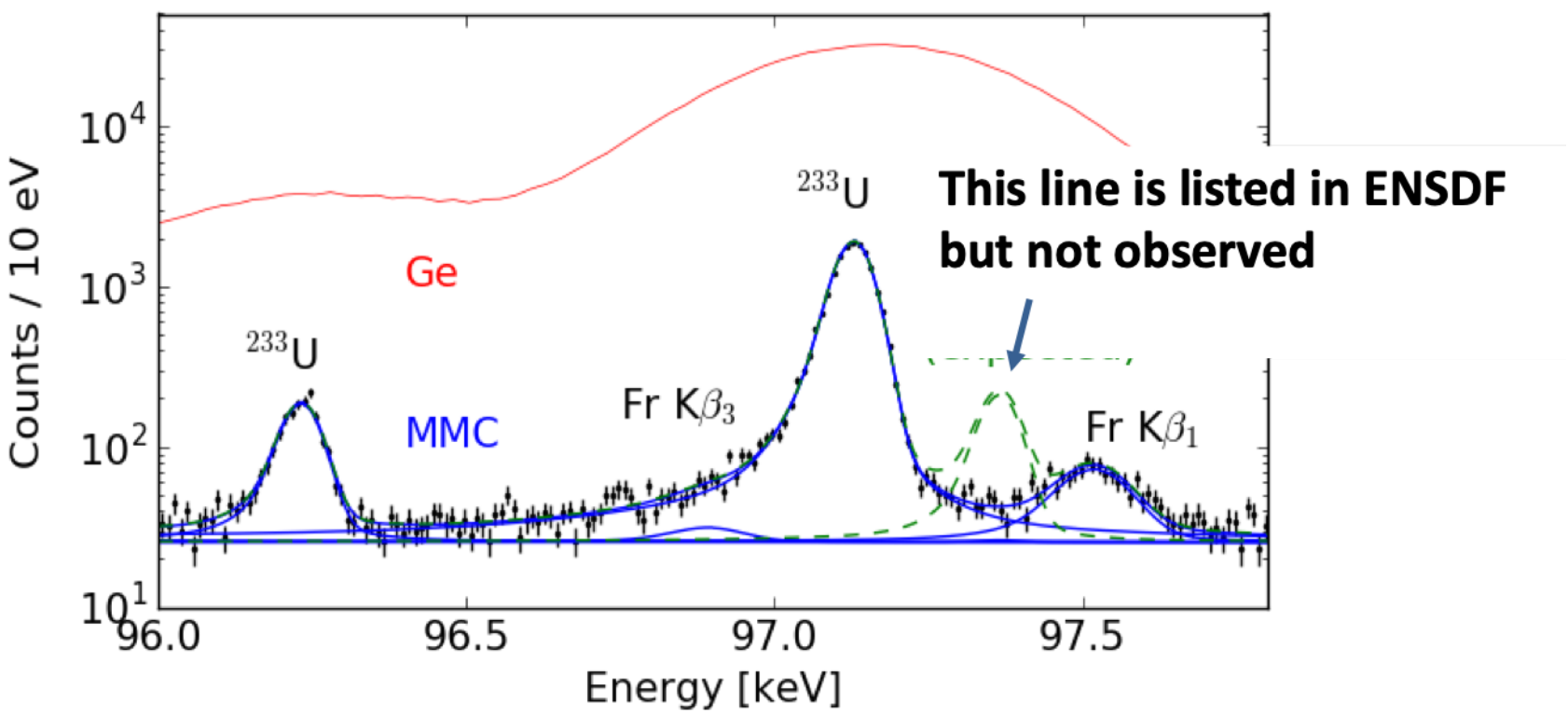

*Figure 2. U-233 gamma-ray spectrum measured by MMC spectrometers. The third peak of U-233 gamma triplet at the 97 keV region is not observed in the MMC spectrum, suggesting that the evaluated energy or emission probability should be reexamined.*

## 2. Microcalorimetry and Nuclear Data (MiND) Workshop

A workshop MiND (Microcalorimetry and Nuclear Data) was organized and sponsored by National Nuclear Security Administration (NNSA) of the Department of Energy (DOE), Office of International Nuclear Safeguards, the U.S. Department of State, Bureau of International Security and Nonproliferation, to respond to growing demands on improved nuclear data as microcalorimetry technology that is now available to end users, such as the International Atomic Energy Agency (IAEA) for high precision radiation spectroscopy and nuclear material analysis.

The goal of the MiND workshop was to identify available microcalorimetry technologies, identify and prioritize nuclear data needs, and provide a roadmap for improving nuclear data, focused on nuclear safeguards applications. The workshop included 53 participants from various microcalorimetry experts and users, nuclear data evaluators and managers, IAEA nuclear material laboratory, and sponsors.

Priority nuclear data has been determined based on workshop discussion and feedback from users and example cases are listed in section 3. To improve priority nuclear data, multi-institutional collaboration has been formed between microcalorimetry experts, users, and nuclear data evaluators, for fast turnaround nuclear data improvement program with round-robin measurement campaign between multiple US DOE laboratories, with sponsorship from US DOE NNSA office of international nuclear safeguards and office of nuclear physics in the DOE. Workshop materials are available at https://conferences.lbl.gov/event/1243/.

## 3. Nuclear Data Needs for Microcalorimetry

A few priority cases for improving nuclear data are listed. Nuclear data listed in this paper is from National Nuclear Data Center [7] and ENSDF library.

### 3.1. Nuclear Data for Uranium Enrichment Measurement

The most precise laboratory method for non-destructive uranium enrichment measurement is using gamma-rays and X-rays in the 90 keV – 100 keV region. The gamma-ray doublet lines at the 92.38(1) keV and 92.80(2) keV from Th-234 decays (daughter isotope of U-238) represent the quantity of U-238, and an X-ray emission line at 93.35 keV (Th $K_{\alpha 1}$ X-rays) from U-235 decays represents the quantity of U-235. Uranium enrichment level can be obtained using their peak areas, emission intensities, and half-lives of U-238 and U-235. The major advantage of using these lines is that those lines are very close in energy, thus variation of detection efficiency is very small and corresponding uncertainty is minimal.

One drawback of current approach is that HPGe detectors cannot resolve the 92 keV doublet. Therefore, nuclear data for the 92 keV doublet, energies and intensities, are still imprecise. Microcalorimeters are promising alternatives for this method as they can fully separate this doublet and can improve accuracy of analysis. Very precise detection efficiency curve at 92 keV – 94 keV range can be obtained using precise measurement of fully separated 92.38(1) keV and 92.80(2) keV lines. For this, improved nuclear data, such as energies and intensities of these two lines, as well as the 93.35 keV X-ray line from U-235 decay should be obtained.

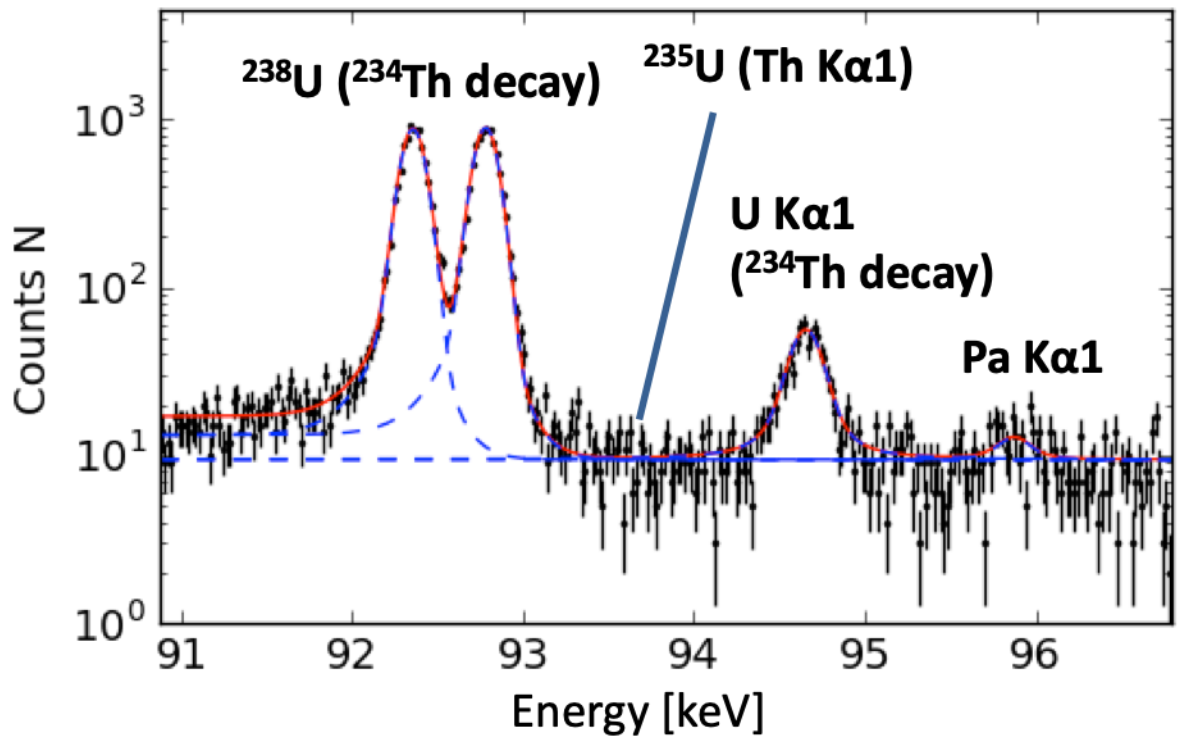

| Parent | Energy [keV] | Intensity [%] |
|---|---|---|
| $^{238}$U ($^{234}$Th) | 92.38(1) | 2.13(20) |
| $^{238}$U ($^{234}$Th) | 92.80(2) | 2.10(20) |
| $^{235}$U (Th Kα1) | 93.35 | 4.7(5) |
| $^{238}$U ($^{234}$Th) | 63.30 | 3.7(4) |

*Figure 3. (Left) Th-234 gamma-ray spectrum measured by MMC gamma-ray spectrometers. 92 keV doublet is separated and quantified for their energies and gamma intensities. Accurate values of their intensities and energies can significantly improve the accuracy of uranium enrichment analysis using microcalorimetry. (Right) Priority gamma-ray and X-ray lines for 90 keV region analysis using microcalorimetry.*

3.2. Direct gamma emission of U-238

The major drawback for the uranium enrichment measurement technique with 92 keV is that it assumes secular equilibrium of uranium sample and thus it cannot be applied for samples with recent chemical separation. 92 keV doublet is from the $^{234}$Th daughter isotope with 24.10(3) days half-life, thus samples less than several months from chemical separation cannot be measured by this technique. However, the 92 keV from daughter nuclei has been the most precise non-destructive assay method because U-238 itself does not have strong gamma-ray emissions. There are two weak gamma lines at 49.55(6) keV and 113.5(1) keV with intensities of 0.064(8)% and 0.0102(15)% respectively, but they are difficult to detect by current technologies. The 49.55(6) keV line is detectable by HPGe detectors, however The nearby 49.46-keV line from U-236 interferes with this measurement and prevent using the 49.55(6) keV for direct quantification of U-238.

The 113.50(10)-keV line may also be useful for direct U-238 quantification. Microcalorimeters can detect this line above background, however its intensity (currently at 15% relative uncertainty) should be determined more precisely.

Figure 4 shows an experimental HPGe spectrum and a simulated microcalorimeter gamma spectrum of a 10% enriched uranium sample. Microcalorimeters with excellent energy resolution may be able to utilize this 49.55 keV line as well as the 113.5(1) keV line of U-238. For this, energies as well as intensities of 49.55(6) keV from U-238 decay and 49.46(10) keV from U-236 decay should be improved, because their relative energy difference is 90 eV $\pm$ 116 eV and therefore poorly constrained. Even microcalorimeters may not be able to resolve these lines if their energy difference is smaller than the energy resolution of microcalorimeters.

Also, gamma-ray lines from U-234 and U-235, at 53.20(2) keV (U-234 decay), 51.21(5) keV (U-235 decay), and 54.25(5) keV (U-235 decay) need to be improved as they can be used for another new method of uranium enrichment measurement at 50 keV region.

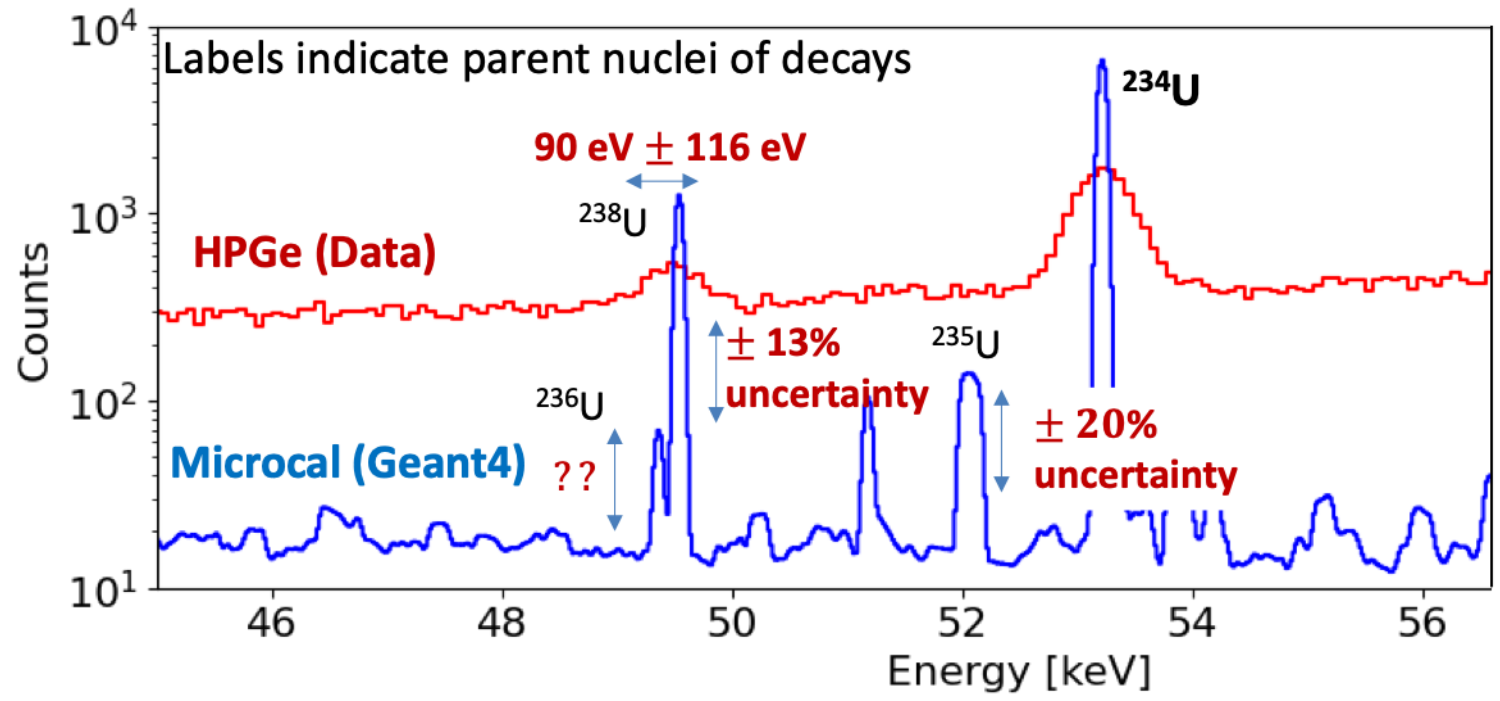

| Parent | Energy [keV] | Intensity [%] |
|---|---|---|
| $^{238}$U | 49.55(6) | 0.064(8) |
| **$^{236}$U** | **49.46(*10*)** | **0.078(12)** |
| $^{232}$U | 57.78(5) | 0.1999(18) |
| $^{234}$U | 53.20(2) | 0.1230(20) |
| $^{235}$U | 51.21(5) | 0.034(7) |
| $^{235}$U | 54.25(5) | 0.015(15) |
| $^{238}$U | 113.50(10) | 0.0102(15) |

*Figure 4. (Left) An experimental HPGe spectrum and a simulated MMC spectrum of 10% enriched uranium sample. The U-236 interference is the major challenge for direct quantification of U-238 using the 49.55(6) keV line. Microcalorimeters could resolve these overlapping lines, if their energy difference is larger than the energy resolution of microcalorimeters. However, energies and intensities of the 49 keV lines are very poorly known. (Right) List of uranium gamma-rays that need to be improved for direct $^{238}$U quantification and isotopic composition analysis.*

## 3.3. Nuclear Data for Plutonium Assay

TES gamma-ray spectrometer systems (SOFIA, [5]) have been deployed at laboratories [5] for high resolution gamma-ray spectroscopy. One of the primary applications is high precision plutonium analysis in the 100 keV – 250 keV regions. One of primary uncertainty factors of HPGe-based Pu assay was uncertainties of line energies and X-ray widths, as many of gamma-ray and X-ray lines were overlapped in such low energy region and spectral fitting of overlapping lines relied on known energies and X-ray widths of overlaps [9]. Microcalorimeters can resolve complex gamma-ray and X-ray lines, thus uncertainties arising from line energies and X-ray widths could be significantly reduced [9]. Remaining dominant uncertainty factor is intensities of gamma-ray and X-ray lines. Especially, TES microcalorimetry uses "anchor" gamma-ray lines (Table 1) for calibrating detection efficiency curve. Intensities of those anchor lines significantly affect detection efficiency curve and thus accuracy of Pu isotopic analysis. These uncertainties range from 0.5% to 1.0% and they are the primary uncertainty factors in microcalorimetry analysis. Improving uncertainties of these anchor lines is one of the top priorities for microcalorimetry.

*Table 1. (Left) Anchor gamma-ray lines that are used for plutonium gamma spectra analysis using microcalorimeters. (Right) Pu lines being used for isotopic ratio analysis. Uncertainties of intensities of these lines directly affect the accuracy of Pu analysis.*

Anchor lines

| Parent | Energy [keV] | Intensity [%] |
|---|---|---|
| $^{239}$Pu | 129.296(1) | 6.31(4)E-3 |
| $^{241}$Pu | 148.567(10) | 1.860(14)E-4 |
| $^{240}$Pu | 160.308(3) | 4.058(15)E-4 |
| $^{239}$Pu | 195.679(8) | 1.070(10)E-4 |
| $^{239}$Pu | 203.550(5) | 5.69(3)E-4 |

Analysis lines

| Parent | Energy [keV] | Intensity [%] |
|---|---|---|
| $^{238}$Pu | 99.853(3) | 7.29(8)E-3 |
| $^{239}$Pu | 98.780(20) | 1.47(7)E-3 |
| $^{239}$Pu | 125.21(10) | 5.63(15)E-5 |
| $^{240}$Pu | 104.237(4) | 7.14(8)E-3 |
| $^{241}$Pu | 159.955(?) | 6.54(8)E-6 |
| $^{241}$Pu ($^{237}$U) | 101.059 | 24.5(6) |
| $^{241}$Pu ($^{237}$U) | 208.005(23) | 21.2(3) |

## 4. Improving Nuclear Data with Microcalorimeters

A large set of current nuclear decay data is relatively accurate, as they have been measured and evaluated through many decades. Improving nuclear data to beyond current accuracy can largely benefit from a new technology with superior energy resolution, like microcalorimeters. Microcalorimeters have an order of magnitude better energy resolution and thus can dramatically improve precision of measurement. Nuclear data measurements require very accurate calibration of detector responses, such as detection efficiency, peak shapes, energy resolution, and energy calibration. Also, understanding systematic uncertainties, particularly those specific in each measurement system, should be well understood to define any underlying biases or uncertainties of measurement results.

One of the best-known calibration sources at low energy (a few 100 keVs) region is Yb-169. Yb-169 has energy values measured by double crystal spectrometers thus independently and precisely measured with very high accuracy. Its gamma-ray intensities are measured by large number of independent experiments and thus the best candidate for accurate calibration of microcalorimeters and HPGe detectors as well for the low energy gamma-rays.

The MiND workshop participants formed a collaboration to improve the priority nuclear data for international nuclear safeguards, using various microcalorimeter and conventional detector technologies. Idaho National Laboratory (INL), Lawrence Berkeley National Laboratory (LBNL), Lawrence Livermore National Laboratory (LLNL), Los Alamos National Laboratory (LANL), National Institute of Standards and Technology (NIST) – Boulder, Pacific Northwest National Laboratory (PNNL), University of California – Berkeley (UCB), and University of Colorado (CU) (listed in alphabetic order), have started reviewing current nuclear data library and literature of priority nuclear data, conducting preliminary measurements with microcalorimeters for designing round-robin measurement campaign with standard uranium and plutonium sources mixed with Yb-169. This collaboration includes nuclear physicists to produce Yb-169 source materials by nuclear reactions, radiochemists for purifying Yb-169, U, and Pu sources with mass-spectrometry for quantifying source compositions, and microcalorimetry experts for conducting nuclear data measurements, and nuclear data evaluators for reviewing current nuclear data, advising experimental plans, and reviewing measurement results.

## 5. Conclusion

Technical readiness levels of microcalorimetry technology have been significantly advanced in the past decade, and microcalorimeters are now deployable to end-users. Microcalorimeters have a strong potential to improve analysis accuracy by an order of magnitude if corresponding nuclear data is improved. Scientists from various backgrounds met at the MiND workshop to discuss nuclear data needs, priorities, and roadmap for improving them. A set of priority nuclear data for microcalorimeter gamma-ray spectrometry has been identified. A multi-institutional collaboration has been formed to improve the priority nuclear data via round-robin measurement campaign.

**Acknowledgments:** This work was performed under the auspices of the U.S. Department of Energy by Lawrence Livermore National Laboratory under contract DE-AC52-07NA27344, by Lawrence Berkeley National Laboratory under contract DE-AC02-05CH11231, and by Los Alamos National Laboratory under contract 89233218CNA000001. The work presented in this paper was funded by the National Nuclear Security Administration of the Department of Energy, Office of International Nuclear Safeguards, the U.S. Department of State, Bureau of International Security and Nonproliferation.